# REPORTS

# Demonstration of a Metamaterial Electromagnetic Cloak at Microwave Frequencies

D. Schurig,[1] J. J. Mock,[1] B. J. Justice,[1] S. A. Cummer,[1] J. B. Pendry,[2] A. F. Starr,[3] D. R. Smith[1]*

Combining the tools for transforming space-time developed for General Relativity with the capabilities of artificially structured metamaterials, an entirely new means of controlling electromagnetic fields has emerged. Here, we utilize a coordinate transformation in which a hole is opened up in space. The transformation provides a complete prescription for an electromagnetic cloak which, although complex, can be readily constructed with metamaterials.

A new approach to the design of electromagnetic structures has recently been proposed, in which the trajectories of electromagnetic waves are controlled by materials that mimic a distortion of space (1, 2). Although it is not feasible to distort space or space-time in arbitrary ways (which would require complex distributions of positive and negative mass), the electromagnetic behaviour of any space can be realized by tailoring the electromagnetic response properties of a medium. In this way, some of the mathematical apparatus developed for General Relativity can be applied to the design of structures that offer new paths to the control of electromagnetic fields, opening up the new paradigm of transformation optics for electromagnetic design (3,19).

While there are many possible applications of transform optics and transform media, a particularly compelling one is that of electromagnetic cloaking, in which a material is used to render a volume effectively invisible to incident radiation. The design process for the cloak is relatively straightforward: we imagine a coordinate transformation that squeezes space from a volume into a shell surrounding the concealment volume. We next note that Maxwell's equations are form invariant to coordinate transformations, only the permittivity, ε, and permeability, μ, are affected by the transformation, becoming both spatially varying and anisotropic. By implementing these complex material properties, the concealed volume plus the cloak appear to have the properties of free space when viewed externally. The cloak thus neither scatters waves nor imparts a shadow in the transmitted field—either of which would enable the cloak to be detected. Other approaches to invisibility either rely on the reduction of backscatter, or make use of a resonance in which the properties of the cloaked object and the cloak itself must be carefully matched (4, 5).

It might be of concern that we are able to achieve two different solutions to Maxwell's equations that both have, in principle, the exact same field distributions on a surface enclosing the region of interest. Indeed, the uniqueness theorem would suggest that these two solutions would be required to have the exact same medium within the surface. The uniqueness theorem, however, applies only to isotropic media (6, 7); the required media that result from our coordinate transformations are generally anisotropic. Such media have been shown to support sets of distinct solutions having identical boundary conditions (8, 9).

The effectiveness of a transformation based cloak design was first confirmed computationally in the geometric optic limit (1,19), and then in full-wave, finite-element simulations (13). Tremendous advances in metamaterials research (10), especially with respect to gradient index lenses (11, 12), have made the physical realization of the specified, complex material properties feasible. In this first such demonstration we have chosen to implement a two dimensional cloak, due to its simpler fabrication and measurement requirements. Recently, we have demonstrated the capability of obtaining detailed spatial maps of the amplitude and phase of the electric field distribution internal to two dimensional, negative index, metamaterial samples at microwave frequencies (14). Using this measurement technique, we confirm the working of our cloak by comparing our measured field maps to simulations.

For the cloak design, we start with a coordinate transformation that compresses space from the cylindrical region $0 < r < b$

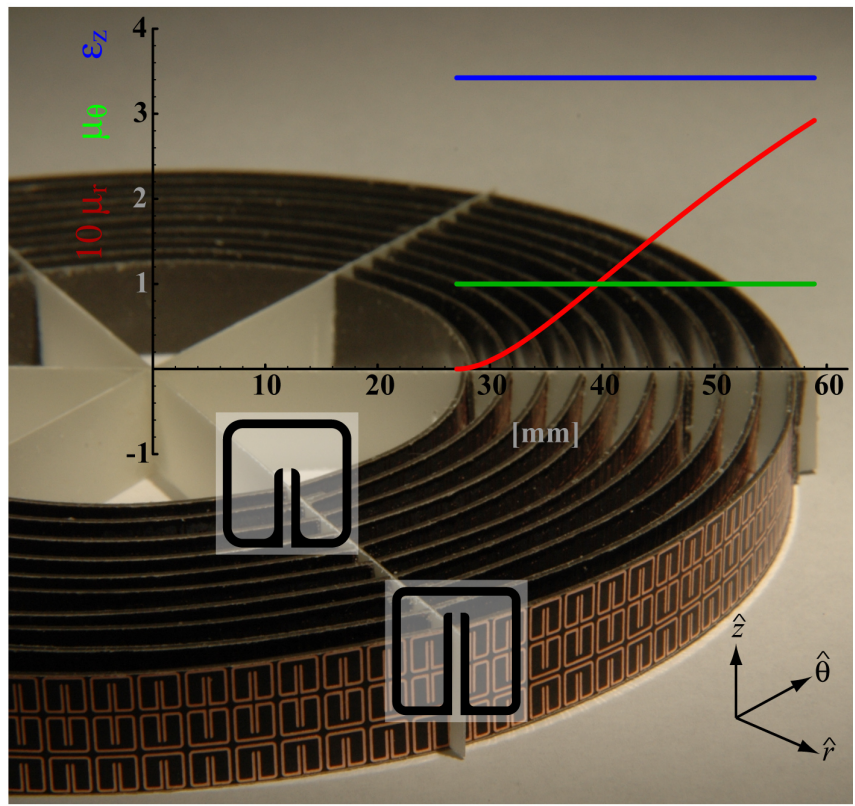

**Fig.1:** Two dimensional microwave cloaking structure with a plot of the material parameters that are implemented. $\mu_r$ (red) is multiplied by the factor 10 for clarity. $\mu_\theta$ (green) has the constant value of 1. $\varepsilon_z$ (blue) has the constant value 3.423. The SRR of cylinder 1 (inner) and cylinder 10 (outer) are shown in expanded schematic form.

[1]Department of Electrical and Computer Engineering, Duke University, Box 90291, Durhman, NC 27708, USA. [2]Department of Physics, The Blacket Laboratory, Imperial College, London SW7 2AZ, UK. [3]SensorMetrix, San Diego, CA, USA.

*To whom correspondence should be addressed. E-mail: drsmith@ee.duke.edu

into the annular region $a < r' < b$. A simple transformation that accomplishes this goal is the following,

$$r' = \frac{b-a}{b}r + a \quad \theta' = \theta \quad z' = z , \quad (1)$$

where the primed and corresponding unprimed variables are the cylindrical position variables in the two coordinate systems. This transformation leads to the following expression for the permittivity and permeability tensor components:

$$\varepsilon_r = \mu_r = \frac{r-a}{r} \quad \varepsilon_\theta = \mu_\theta = \frac{r}{r-a}$$
$$\varepsilon_z = \mu_z = \left(\frac{b}{b-a}\right)^2 \frac{r-a}{r} \quad (2)$$

Details of these calculations are given in the Supporting Online Material (SOM). Eq. 2 shows that all of the tensor components possess gradients as a function of radius, implying a very complicated metamaterial design. However, because of the nature of the experimental apparatus—in which the electric field is polarized along the z-axis (cylinder axis)—we benefit from a significant simplification in that only $\varepsilon_z$, $\mu_r$ and $\mu_\theta$ are relevant. Moreover, if we wish to demonstrate primarily the wave trajectory inside the cloak, which is solely determined by the dispersion relation, we gain even more flexibility in choosing the functional forms for the electromagnetic material parameters. In particular, the following material properties

$$\varepsilon_z = \left(\frac{b}{b-a}\right)^2 \quad \mu_r = \left(\frac{r-a}{r}\right)^2 \quad \mu_\theta = 1 \quad (3)$$

have the same dispersion as those of Eq.(2), implying that waves will have the same dynamics in the medium. In the geometric limit, for example, rays will follow the same paths in the media defined by Eq. (2) or Eq. (3) and refraction angles into or out of the media will be the same *(13)*. The only penalty for using the reduced set of material properties, Eq.(3), is a non-zero reflectance.

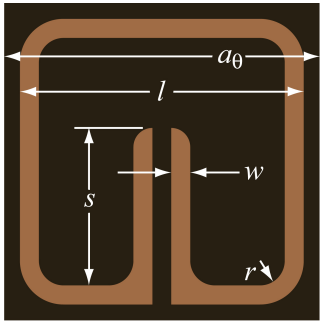

| cyl. | r | s | $\mu_r$ |
|---|---|---|---|
| 1 | 0.260 | 1.654 | 0.003 |
| 2 | 0.254 | 1.677 | 0.023 |
| 3 | 0.245 | 1.718 | 0.052 |
| 4 | 0.230 | 1.771 | 0.085 |
| 5 | 0.208 | 1.825 | 0.120 |
| 6 | 0.190 | 1.886 | 0.154 |
| 7 | 0.173 | 1.951 | 0.188 |
| 8 | 0.148 | 2.027 | 0.220 |
| 9 | 0.129 | 2.110 | 0.250 |
| 10 | 0.116 | 2.199 | 0.279 |

**Fig.2:** SRR design. The in-plane lattice parameters are, $a_\theta = a_z = 10/3$ mm. The ring is square with $l = 3.000$ mm, and trace width, $w = 0.200$ mm. The substrate is 381 μm thick Duroid 5870, $\varepsilon = 2.33$, $t_d = 0.0012$ at 10GHz. The copper film, from which the SRRs are patterned, is 17 μm thick. The parameters, $r$ and $s$, are given in the table together with the associated value of $\mu_r$. The extractions gave roughly constant values for the imaginary parts of the material parameters, 0.002 and 0.006 for the imaginary part of $\varepsilon$ and $\mu$ respectively. The inner cylinder is 1 and the outer is 10.

To implement the material specification in Eq. (3) in a metamaterial, we must design the appropriate unit cells and their layout, which for our implementation represents a pattern that is neither cubic nor even periodic. Both the layout and the unit cells share design parameters, making it advantageous and necessary to optimize them both at once. In this case, Eq. 3 shows that the desired cloak will have constant $\varepsilon_z$ and $\mu_\theta$, with $\mu_r$ varying radially throughout the structure. This parameter set can be achieved in a metamaterial in which split ring resonators (SRRs)—known to provide a magnetic response that can be tailored *(15)*—are positioned with their axes along the radial direction (see Fig. 1).

As can be seen from Eq. 3, the transformed material properties depend strongly on the choice of the cloak inner radius, $a$, and outer radius, $b$. Due to constraints from the unit cell design and layout requirements, we choose the seemingly arbitrary values, $a = 27.1$ mm and $b = 58.9$ mm. The resulting material properties are plotted in Fig.1.

All metamaterials reported to date have consisted of elements repeated in cubic or other standard lattice configurations and diagonal in the Cartesian basis. The layout of our cylindrical cloak, however, uses cells that are diagonal in a cylindrical basis and has "unit cells" that are curved sectors with varied electromagnetic environments. The correct retrieval procedure that would obtain the effective medium properties from such irregular unit cells is not yet available. Given that the curvature is not extreme in this cloak design, however, we model the unit cells as right rectangular prisms in a periodic array of like cells, with the assumption that the actual cells will produce minor corrections in the effective medium properties.

Due to constraints of the layout we chose a rectangular unit cell with dimensions, $a_\theta = a_z = 10/3$ mm and $a_r = 10/\pi$ mm. We are able to obtain both the desired $\varepsilon_z$ and $\mu_r(r)$ from an SRR, by tuning two of its geometric parameters: the length of the split, $s$, and the radius of the corners, $r$, (Fig.2). The parameters, $r$ and $s$, shift the frequency of the electric and magnetic resonance, respectively, though there is some cross-coupling that must be compensated.

Using commercial, full-wave, finite-element simulation software (Microwave Studio, CST), we performed a series of scattering (S-) parameter simulations for SRR unit cells over a discrete set of the parameters $r$ and $s$ covering the range of interest. A standard retrieval procedure *(16)* was then performed to obtain the effective material properties, $\varepsilon_z$ and $\mu_r$, from the S-parameters. The discrete set of simulations and extractions was interpolated to obtain the particular values of the geometric parameters that yielded the desired material properties. We chose an operating frequency of 8.5 GHz which yields a reasonable effective medium parameter, $\lambda/a_\theta > 10$.

The layout consists of ten concentric cylinders, each of which is three unit cells tall. The evenly spaced set of cylinder radii is chosen so that an integral number of unit cells fit exactly around the circumference of each cylinder, necessitating a particular ratio of radial to circumferential unit cell size. We chose to increase the number of unit cells in each successive cylinder by six, enabling us to use six supporting radial spokes that can intersect each of the cylinders in the spaces between the SRRs. This leads to the requirement $a_r/a_\theta = 3/\pi$. Additionally, to minimize the magnetoelectric coupling inherent in single split SRRs *(17)*, we alternated their orientation along the $z$ direction, (Fig.1).

The overall scale of the cloak is such that a complete field mapping of the cloak and its immediate environment is feasible. By the same reasoning, numerical simulations of the cloak, are also feasible, so long as the cloak is approximated by continuous materials. A complete simulation of the actual cloak structure, including the details of the hundreds of SRRs, would be impractical for general optimization studies.

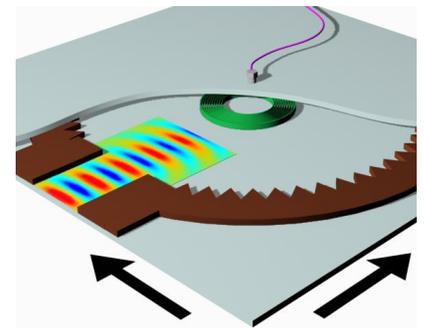

**Fig.3:** Cutaway view of the planar waveguide apparatus. Microwaves are introduced via a coaxial-to-waveguide transition attached to the lower plate and are shaped into an incident beam by walls formed from absorbing material. An antenna mounted in the fixed upper plate measures the phase and amplitude of the electric field. To perform field maps, the lower plate is stepped in the lateral directions.

Simulations of prospective cloaking structures were performed using the COMSOL Multiphysics finite element-based electromagnetics solver, in a manner previously described *(13)*. The performance of a cloak specified by Eq. 2 was evaluated in the context of the step-wise approximation corresponding to the metamaterial cloak (Fig. 1). Other "real-world" effects could also be studied via these simulations, such as the influence of material absorption or slight deviations from the ideal properties, as might be encountered in fabricated samples. Simulated electric field plots comparing the cloaking of structures defined by Eq. 2 and Eq. 3 are shown in Figs. *4A* and *4B*.

For the experimental confirmation, we measure the metamaterial cloak in a parallel-plate waveguide comprising two flat conducting (Al) plates spaced 11mm apart (see Fig. 3). Microwaves are introduced through an X-band (8-12 GHz) coax-to-waveguide adapter that attached to the lower plate. Strips of microwave absorbing material are positioned along either side of the waveguide adapter, forming a channel that launches a microwave beam between the plates. The beam is incident on the cloak, which rests on the lower plate and is nearly of the same height (10 cm) as the plate separation.

A field sensing antenna is formed from a coaxial fixture inserted into a hole drilled through the upper plate. The center conductor of the coaxial connector extends to a position flush with the lower surface of the upper plate and does not protrude into the chamber volume. The lower plate is mounted on two orthogonal linear translation stages, so that the lower plate (including the cloak, waveguide feed and absorber) can be translated with respect to the upper plate (and detector). By stepping the lower plate in small increments and recording the field amplitude and phase at every step, a full, two-dimensional spatial field map of the microwave scattering pattern can be acquired both inside the cloak as well as in the surround free-space region. Further experimental details can be found in *(14)* and in the SOM.

The cloak was placed on the lower plate in the center of the mapping region, and illuminated with microwaves over a discrete set of frequencies that included the expected operating frequency of the cloak. At each frequency, the complex electric field was acquired, and the process repeated for all x- and y- positions in the scan range. After reviewing the field maps at all frequencies, the optimal frequency for the cloak sample was determined to be 8.5 GHz, in near exact agreement with the design target. The best optimal frequency was selected as that which best matched the simulated field plots. The acquired real part of the electric field distribution is shown Fig. *4C*.

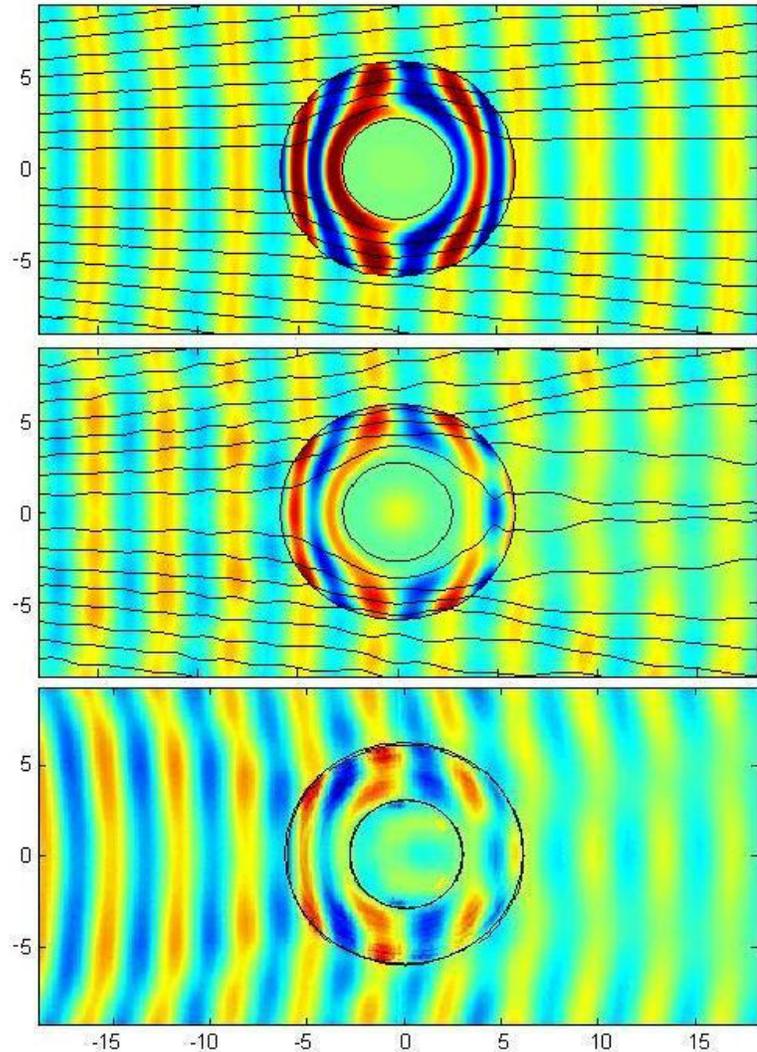

**Fig.4:** Electric displacement field of simulated cloak with exact (A) and reduced (B) material properites. (C) Field amplitude of measured cloak. Black stream lines are tangent to the Poynting vector. Position scales are in centimeter units. Field animations of both the simulations and the data can be viewed in the SOM.

As seen from Fig. *4*, the field plots found using full-wave simulations are in remarkable agreement with the experimental data. The underlying physics of the cloaking mechanism can be studied even further by viewing the field animations (SOM). As the waves propagate through the cloak, a section of the wave front begins to lag as it approaches the inner radius, exhibiting a compression in wavelength and a reduction in intensity. The wave front then separates as it passes through the center of the cloak, eventually "catching up" and rejoining on the opposite side. Notably, the wave fronts at the boundary of the cloak match the wave fronts outside the cloak, which essentially correspond to those of empty space. The scattering is thus minimized, though not perfectly due to the reduced parameter implementation. The fields on the exit side are noticeably reduced due to the absorption of the cloak material.

From an electromagnetic point-of-view, the cloak represents perhaps one of the most elaborate metamaterial structures yet designed and produced. The remarkable agreement between simulation and experiment is evidence that metamaterials can indeed be designed to specific and exacting specifications, including gradients and with non-rectangular geometry. The agreement further demonstrates that fields can be measured accurately inside a two dimensional metamaterial with high resolution.

Though the invisibility demonstrated here is imperfect due to the approximations used and material absorption, this first demonstration of a metamaterial cloak does show the basic physics involved and demonstrates the feasibility of implementing media specified by the transformation method with metamaterial technology.

**References and Notes**
1. J. B. Pendry, D. Schurig, D. R. Smith, *Science* **312**, 1780 (2006).
2. U. Leonhardt, *Science* **312**, 1777 (2006).
3. U. Leonhardt, T. G. Philbin, "General relativity in electrical engineering," arXiv:physics/0607418.
4. A. Alu, N. Engheta, *Phys. Rev. E* **72**, 016623 (2005).
5. G. W. Milton, N.-A. P. Nicorovici, *Proc. Roy. Soc. London A* **462**, 1364 (2006).